\documentclass{mn2e}
\usepackage{graphicx}

\newif\ifAMStwofonts

\input epsf

\def\lesssim{\mathrel{\hbox{\rlap{\hbox{\lower4pt\hbox{$\sim$}}}\hbox{$<$}}}}

\def\gtrsim{\mathrel{\hbox{\rlap{\hbox{\lower4pt\hbox{$\sim$}}}\hbox{$>$}}}}

\def\msun{$M_{\odot}$}

\def\ll_lsun{$\Log({L/\rm L_{\odot}})$~}

\def\masa_msun{$M/ \rm M_{\odot}$~}

\def\m_mstar{$M/M_{*}$~}

\title[Evolutionary and pulsational properties of low-mass carbon-oxygen white
dwarf stars]{Evolutionary and pulsational properties of low-mass white
dwarf stars with oxygen cores resulting from close binary evolution}

\author[L. G. Althaus, A. H. C\'orsico, A. Gautschy, Z. Han,  
A. M. Serenelli and J. A. Panei]
{L.   G. Althaus$^1$
\thanks{Member of the Carrera del Investigador
Cient\'{\i}fico y Tecnol\'ogico, CONICET, Argentina. 
Email: althaus@fcaglp.unlp.edu.ar}, 
A. H. C\'orsico$^1$
\thanks{Fellow of CONICET, Argentina.
Email: acorsico@fcaglp.unlp.edu.ar},
A. Gautschy$^2$,
Z. Han$^3$
\thanks{Email: zhanwen@public.km.yn.cn}, 
A. M. Serenelli$^1$
\thanks{Fellow of CONICET, Argentina.
Email: aldos@MPA-Garching.MPG.DE} and
\newauthor
J. A. Panei$^1$
\thanks{Fellow of CONICET, Argentina.
Email: panei@fcaglp.unlp.edu.ar}\\ 
$^1$Facultad  de  Ciencias
Astron\'omicas y Geof\'{\i}sicas de la Universidad Nacional de La
Plata and \\
Instituto de Astrof\'{\i}sica de La Plata (CONICET-UNLP), Paseo del Bosque 
S/N, (1900) La Plata, Argentina \\ 
$^2$Froburgstr.43, 4052 Basel, Switzerland \\
$^3$National Astronomical Observatories / Yunnan Observatory, The Chinese 
Academy of Science, P.O.Box 110, Kunming, 650011, China\\ }

\date{August 29}

\pagerange{\pageref{firstpage}--\pageref{lastpage}}

\pubyear{2003}

\begin{document}

\maketitle

\label{firstpage}

\begin{abstract} The present work is designed to explore
the evolutionary  and pulsational properties of  low-mass white dwarfs
with carbon/oxygen cores.  In particular, we follow the evolution of a
0.33-\msun white  dwarf remnant  in  a self-consistent  way with  the
predictions of  nuclear burning, element diffusion and  the history of
the white dwarf progenitor. Attention  is focused on the occurrence of
hydrogen shell  flashes induced by diffusion  processes during cooling
phases. The evolutionary stages prior to the white dwarf formation are
also  fully  accounted  for   by  computing  the  conservative  binary
evolution  of an  initially 2.5-\msun  Pop.  I  star with  a 1.25  \msun 
companion,  and  period $P_i$=  3  days. Evolution  is
followed down  to the domain of the  ZZ Ceti stars on  the white dwarf
cooling branch.

We  find  that chemical diffusion 
induces the occurrence of an additional hydrogen  thermonuclear  flash 
which leads  to stellar models  with  thin  hydrogen  envelopes. As a
result, a  fast  cooling is  encountered  at
advanced  stages of  evolution. In addition,  we  explore  the
adiabatic pulsational  properties of the resulting  white dwarf models.
As compared  with their helium-core counterparts, low-mass oxygen-core white 
dwarfs are characterized by a pulsational spectrum much more featured, 
an aspect which could eventually be used for distinguishing both types of 
stars if low-mass white dwarfs were in fact found to pulsate as ZZ 
Ceti~--~type variables. Finally, we perform a non-adiabatic pulsational 
analysis 
on the resulting carbon/oxygen low-mass white dwarf models.  

\end{abstract}


\section{Introduction} \label{sec:intro}

Low-mass white dwarf (WD) stars, having masses less than 
$\approx$ 0.5 \msun, are the result of evolution of certain
close binary systems (Paczy\'nski 1976; Iben \& Webbink 1989; Iben \&
Livio 1993).  Indeed, mass-transfer episodes in binary systems are
required to form low-mass WDs. An isolated star of comparable low mass
would take more than a Hubble time to evolve into a corresponding WD
configuration. The low-mass WDs are thought to have a core composed by
helium; their binary nature was first put on a firm observational
basis by Marsh (1995) and Marsh, Dhillon \& Duck (1995).  Since then,
low-mass WDs have been detected in numerous binary configurations
containing usually either another WD or a neutron star (see, e.g.,
Lundgren et al.  1996; Moran, Marsh \& Bragaglia 1997; Orosz et al.
1999; Maxted et al.  2000, van Kerkwijk et al.  2000).  In addition,
several low-mass WDs have been found in open and globular clusters
(Landsman et al.  1997; Edmonds et al.  1999, 2001; Taylor et al.
2001).  Evolutionary models for helium-core WDs have been calculated
by Hansen \& Phinney (1998), Driebe et
al. (1998); Sarna, Ergma \& Antipova (2000), Althaus, Serenelli \&
Benvenuto (2001a,b) and Serenelli et al. (2002).

However, theoretical  evidence casts  some doubt  on  that all
low-mass WDs be indeed helium-core WDs. In fact, Iben \& Tutukov (1985)
proposed several scenarios where low-mass white dwarfs could
harbor cores with elements heavier than helium. More recently,
Han, Tout \& Eggleton
(2000) have carried out new close binary evolutionary calculations and
found that  some of the presumed  helium-core WDs in double degenerate 
systems may  actually be WDs
with carbon-oxygen cores.   Specifically, their calculations show that,
if the onset  of mass transfer episodes takes  place after the end
of central hydrogen burning during the middle Hertzsprung-gap stage, a
carbon-oxygen WD  with mass as  low as 0.33  \msun may be  formed from
stable Roche lobe overflow (RLOF)  if the initial mass of  the primary 
star is  close to 2.5 \msun.

The present work aims at exploring the evolution of low-mass WDs with
carbon/oxygen cores in a self-consistent way with the predictions of
nuclear burning, time-dependent
element diffusion and the history of the white dwarf progenitor. We
concentrate on a carbon/oxygen-core WD remnant of mass 0.33 \msun.
Attention is focused on the occurrence of hydrogen shell flashes
and the role played by diffusion processes during cooling phases.  The
evolutionary stages prior to the white dwarf formation are also fully
accounted for by computing the conservative binary evolution of an
initially 2.5-\msun Pop.~I star with a 1.25-\msun
companion, and period $P_i$= 3 days.  Finally, we explore the
adiabatic and non-adiabatic pulsational properties of the resulting WD models
and we compare them with those of helium-core WDs. 

In this work, we hope to address essentially the two following
questions: (i) What is the role of element diffusion in the evolution
of low-mass, carbon/oxygen WDs?; and (ii) What differences in the
evolutionary properties as well as in the pulsational signal (if these
objects do actually pulsate) can we expect between low-mass WDs with
helium cores and those with carbon/oxygen cores?  In Section~2, we
shortly describe the main physical inputs of the models.  In
section~3, we present the evolutionary results, particularly regarding
the WD stages.  Section~4 contains the discussion of the pulsational
properties of our models. Finally, Section~5 is devoted to summarizing
our results and making some concluding remarks.

\section{Input physics and computational details}

The  evolutionary  code used  in  this  work  to simulate  the  binary
evolution  is that  described  by  Han et  al.  (2000) and  references
therein. The code  uses a self-adaptive non-lagrangian mesh  and it is
based  on an up-to-date  physical description  such as  OPAL radiative
(Rogers \&  Iglesias 1992) and molecular (Alexander  \& Ferguson 1994)
opacities,  a  detailed  equation   of  state  that  includes  coulomb
interactions and  pressure ionization and nuclear  reaction rates from
Caughlan  et al.  (1985) and  Caughlan \&  Fowler  (1988).  Convective
overshooting is not considered.

For the WD regime, we  employed the LPCODE evolutionary code which has 
been used in previous works on WD evolution. The code has been developed at La Plata Observatory and it is described in Althaus (1996), Althaus  et 
al. (2001a), Serenelli (2002), Althaus et al. (2003)  and references therein.  
As for the constitutive physics, LPCODE employs OPAL radiative opacities 
(including carbon and oxygen-rich compositions)  for arbitrary metallicity  
from Iglesias \& Rogers  (1996).   Opacities  for  various metallicities  
are  required during the WD cooling regime in view of the metallicity 
gradients that develop  as  a result  of  gravitational  settling  
(see below).  The
equation of state is an updated version of that of Magni \& Mazzitelli
(1979).  High-density conductive  opacities and the various mechanisms
of  neutrinos  emission   are  taken  from  the  works   of  Itoh  and
collaborators. Hydrogen burning is taken into account by considering a
complete network of thermonuclear  reaction rates corresponding to the
proton-proton chain and the  CNO bi-cycle.  Nuclear reaction rates are
taken from Caughlan \& Fowler (1988). Electron screening is treated as
in Wallace, Woosley \& Weaver (1982).

An important aspect of the present study is the evolution of the
chemical abundance distribution caused by element diffusion during the
whole WD stage.  In our treatment of time-dependent diffusion we have
considered the following nuclear species: $^{1}$H, $^{3}$He, $^{4}$He,
$^{12}$C, $^{14}$N and $^{16}$O. The chemical evolution resulting from
element diffusion is described in LPCODE, for a given isotope $i$ having a
number density $n_i$, by the continuity equation as

\begin{equation}
\frac{\partial n_i}{\partial t}= -\frac{1}{r^2}
\frac{\partial}{\partial r} \left(r^2 n_i w_i
\right), 
\end{equation}

\noindent where  $w_i$  is  the  diffusion velocity as  given by  the
solution of the multicomponent flow equations describing gravitational
settling, chemical and thermal diffusion (Burgers 1969). Because we are
interested in the chemical evolution occurring quite deep in the star,
radiative  levitation  (which  is   expected  to  modify  the  surface
composition of hot  WDs) has been neglected. In  terms of the gradient
of ion densities, diffusion velocities can be written in the form

\begin{equation}
w_i = w_{i}^{\rm gt} - \sum\limits_{{\rm ions} (j)} \sigma_{ij}\
\frac{d\ln{n_j}}{dr} ,
\end{equation}

\noindent where $w_{i}^{\rm gt}$ stands for the velocity component due
to gravitational settling and  thermal diffusion and $\sigma_{ij}$
represent the components due to the gradients in number density
(chemical diffusion). 
Details are given in
Althaus  et al. (2001a) and  Gautschy \&  Althaus (2002).  We want  to 
mention that  after computing the
change  of  abundances  by  effect  of  diffusion,  they  are  evolved
according  to nuclear reactions  and convective  mixing.  It  is worth
mentioning that  radiative opacities are  calculated for metallicities
consistent with diffusion  predictions. In particular, the metallicity
is taken as two times the abundances of CNO elements.

In this work, we have followed the {\it complete} evolution of an
initially 2.5-\msun star in a close binary system from the main
sequence through the mass transfer episodes to the domain of the ZZ
Ceti stars on the WD cooling branch.  Specifically, the initial mass
of the primary and secondary stars is, respectively,  2.5
and 1.25 \msun with an initial orbital period of
$P_i$= 3 days. A solar-like initial composition (Y,Z)= (0.28, 0.02)
was adopted.  The treatment of convection is that of the mixing-length
theory with the ratio of the mixing length to the local pressure
scale-height set to two (see Han et al.  2000 for details). The onset
of RLOF, which is assumed to be conservative, occurs after the central
hydrogen exhaustion in the middle of the Hertzsprung-gap stage.  After
mass transfer episodes, the mass of the WD remnant is 0.33 \msun. 
As stated in the introduction,
one of the aim of this investigation is the exploration of the
influence that diffusion has on the evolution of low-mass,
carbon/oxygen WDs.  Thus, in addition to evolutionary models in which
diffusion is considered, we have computed the WD evolution for the
case when diffusion is neglected.  This allowed us to clearly
identify the effect induced by diffusion on the evolution of such
objects.

\section{Evolutionary results} \label{sec:results}

\begin{table*}
\centering
\begin{minipage}{120mm}
\caption{Selected stages for 0.33-\msun \ oxygen-core WD models considering 
element diffusion, as labeled in Fig. \ref{fighrdif}. From left to right: 
surface luminosity,
effective temperature,  age, surface
gravity,  nuclear  luminosity, surface hydrogen abundance
and hydrogen envelope mass.}
\begin{tabular}{@{}cccccccc@{}}
\hline
$$ & Log(${L/\rm L_{\odot}}$)  &
Log($T_{\rm eff}$) & Age(10$^6$ yr) & Log($g$) &
Log($L_{\rm nuc}/{\rm L_{\odot}})$ & $X_H$ & Log($M_{\rm H}/{\rm M_{\odot}})$ \\
\hline
(A) & 1.2715 & 4.5640 & 0.000 & 5.8952 & 2.843 & 0.573 & -2.9003 \\
(B) & -0.4906 & 4.2337 & 7.797007 & 6.3360 & 5.218 & 1.000 & -3.4008 \\ 
(C) &-1.4003 & 3.8962  & 7.797009& 5.8957 & 4.657 & 1.000 & -3.4008 \\ 
(D) & 3.4317 & 3.7982  & 7.797042 & 0.6717 & 3.2980 & 0.304 & -3.4008 \\ 
(E) & 2.5554 & 4.9168 & 7.81 & 6.0225 & 2.544 & 0.476 & -3.5674 \\
(F) & 0.0000 & 4.5066 & 7.98 & 6.9369 & -1.499 & 1.000 & -3.5987 \\
(G) &-1.9967 & 4.100 & 184.2  & 7.3095 & -2.968 & 1.000 & -3.6502\\
(H) &-4.0289 & 3.635 & 4085  & 7.4791 & -6.982 & 1.000 & -3.6567\\
\hline

\end{tabular}
\medskip
\vskip -0.35cm
{\small Ages are counted from point A, which corresponds to an evolutionary
stage just before the onset of the first thermonuclear flash. The surface hydrogen
abundance is evaluated at  a
mass depth of $10^{-10} M_*$ below the stellar surface. } 
\end{minipage}

\end{table*}

\begin{figure}
\includegraphics[width=8cm,height=8cm,angle=-90]{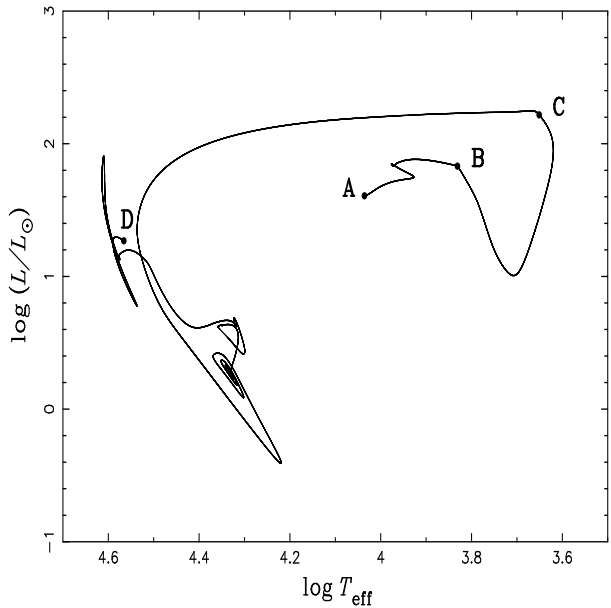}
\caption{Hertzsprung-Russell diagram for the
primary star during the binary evolution stage and core helium burning 
stages. Labels along the track denote specific stages during the evolution 
and are discussed in the text.}
\label{hrd}
\end{figure}

In the following, we describe the
evolution on the  Hertzsprung-Russell diagram (HRD) of the WD progenitor. 
Specifically, in Fig.\ref{hrd} we show the HRD of the primary star, 
which initially has a stellar mass of 2.5 \msun. 
The primary evolves from zero-age main sequence (point A) and burns 
hydrogen at its centre. After central hydrogen is exhausted, the star 
evolves to the Hertzsprung gap and fills its Roche lobe at point B, 
and the RLOF phase begins. 
The envelope mass is transferred to the secondary, which initially has a stellar mass of 1.25 \msun, till RLOF finally
stops at point C. After the RLOF, the primary becomes a helium star (with
a thin hydrogen layer). As a result of mass loss episodes, the stellar 
mass has decreased to $M_{\rm 1f}= 0.3293$ \msun, 
while the mass of the secondary  is now  $M_{\rm 2f}= 3.4207$ \msun 
and the orbital period has increased to $P_{\rm f}= 64.08\ {\rm d}$.
The helium star contracts and moves quickly across the HRD. Helium
is ignited at the centre and the star experiences ``breathing pulses''
(the loops at lower-left corner of the figure; see e.g.
Castellani et al.\ 1985, Han et al.\ 2002). After central helium-burning, the
star has an almost pure oxygen core of $0.11$ \msun. The evolution 
calculation of the resulting oxygen-core WD is continued from point D.

The resulting WD binary system continues to evolve and the secondary
will fill its Roche lobe as an asymptotic giant branch (AGB) star and
leads to the onset of another RLOF.
The star has a CO core and a deep convective envelope. The mass transfer
is dynamically unstable and a common envelope (CE: Paczy\'nski, 1976)
is formed. The CE engulfs the CO core and the primary (a WD). A large
amount of orbital energy is deposited into the CE due to
the friction between the CE and the embedded binary. The CE may
be ejected if the orbital energy released can overcome the binding energy of
the envelope, and leads to the formation of a binary system with
two WD components, i.e. a double degenerate. If the
common envelope ejection efficiency is $\alpha_{\rm CE}= 1$ and
thermal contribution to CE ejection is $\alpha_{\rm th}= 1$ 
(see Han 1998; Han et al.\ 2002 for the definitions), 
the double degenerate system
will have $M_1= 0.3293$ \msun, $M_2=0.6572$ \msun, $P=0.1837\ {\rm d}$.
However, if $\alpha_{\rm CE}= \alpha_{\rm th}=0.75$,
the orbital period of the double degenerate will be
0.0735 d instead.

\begin{figure}
\includegraphics[width=8cm,height=8cm]{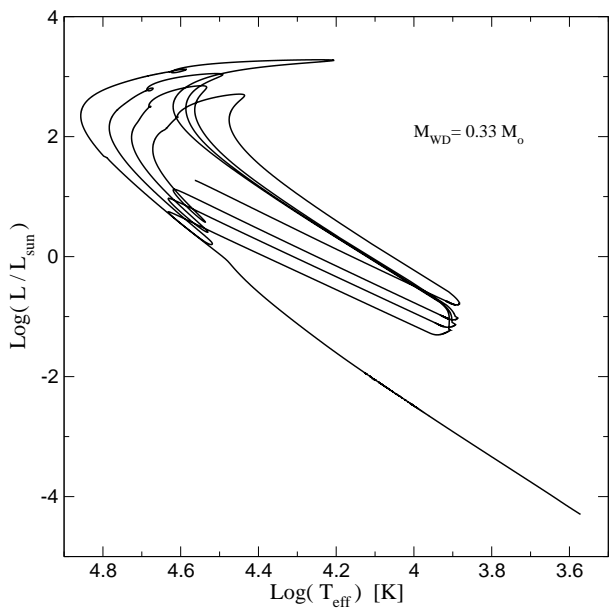}
\caption{Hertzsprung-Russell diagram for the WD evolution of the 
0.33-\msun oxygen-core remnant when element diffusion is negelected.
The WD experiences four hydrogen-shell flash episodes before reaching the 
final cooling branch.}
\label{fighrsdif}
\end{figure}

\begin{figure}
\includegraphics[width=8cm,height=8cm]{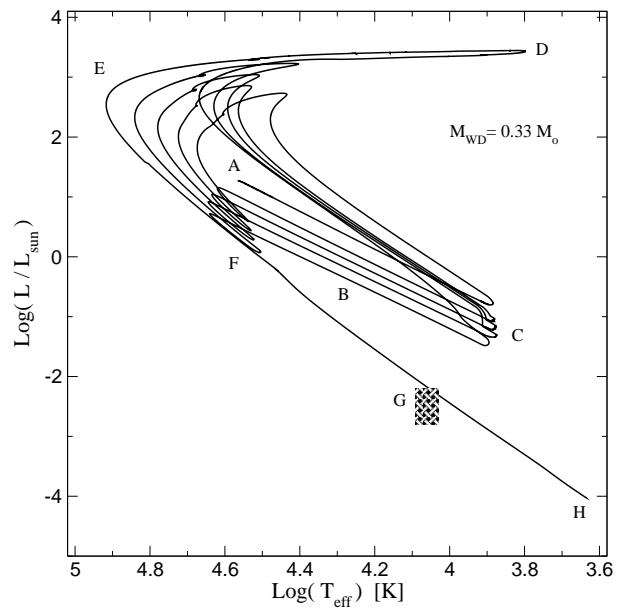}
\caption{Same as Fig.~\ref{fighrsdif} but for the case when element 
diffusion is considered. Letters at various positions along the track 
indicate selected stages of
the evolution described in Table 1. Also, the approximate location of
the domain of ZZ Ceti instability strip is shown as a small shaded region. 
As a result of hydrogen chemically diffusing inwards during the cooling track, 
the WD experiences an additional thermonuclear flash, as compared with the   
situation  of no diffusion. As a result of this diffusion-induced flash, the 
star for a brief time interval, is forced to evolve back to the red giant 
domain (point D).}
\label{fighrdif}
\end{figure}

In the following, we describe the main evolutionary results for the WD
evolution   of   the   0.33-\msun   oxygen-core  remnant.   In   Figs.
\ref{fighrsdif} and \ref{fighrdif} we show  the HRD for the WD remnant
when element diffusion is  neglected and when diffusion is considered,
respectively.   Labels   along  the  track  from  A   to  H  represent
evolutionary   stages   for    models   with   diffusion,   the   main
characteristics  of which  are  listed  in Table  1.   In both  cases,
evolution is  characterized by the  occurrence of several  episodes of
thermonuclear  flashes  during which  the  star experiences  extensive
loops in the  Hertzsprung-Russell diagram. Each of these  loops is due
to  unstable  nuclear  burning  at  the bottom  of  the  hydrogen-rich
envelope.   Note   that  when  diffusion  is   considered,  the  model
experiences an  additional thermonuclear  flash, as compared  with the
situation of  no diffusion. This  last flash is triggered  by chemical
diffusion which, as  the WD evolves along the  cooling branch, carries
some hydrogen inwards to hotter  layers, where it burns unstably\footnote
{ The
occurrence   of  diffusion-induced   hydrogen  shell   flashes   is  a
theoretically known  phenomena  found  in some detailed 
white dwarf evolutionary
calculations (see for instance, Iben \& MacDonald
1986 in  the  context  of intermediate-mass,  carbon-oxygen  white
dwarfs).}.  As  we shall see,  the occurrence of  this diffusion-induced
thermonuclear flash is critical  regarding the subsequent evolution of
the remnant  even during its final  cooling phase. It  is worth noting
that the  diffusion-induced flash is the  most energetic one  and as a
result the star,  for a brief time interval,  reaches giant dimensions
again  (point D).  After this  flash episode,  the  remnant eventually
settles on its terminal cooling track.

\begin{figure}
\includegraphics[width=8cm,height=8cm]{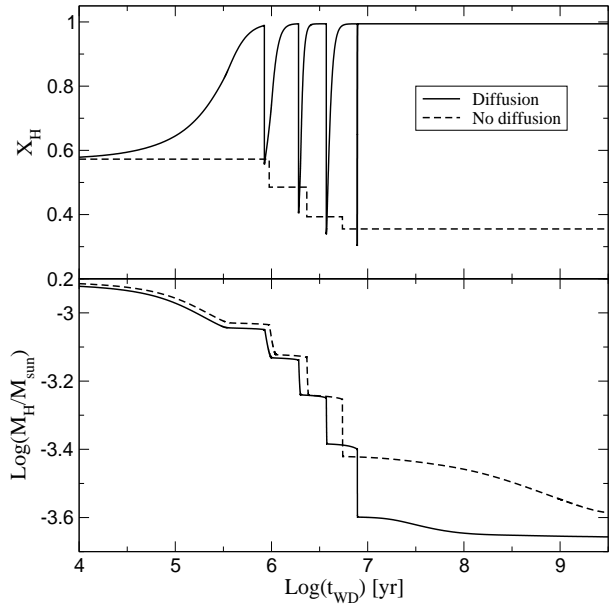}
\caption{Abundance by mass of hydrogen at a mass depth of $10^{-10}
M_*$ below the stellar surface (upper panel) and mass of the hydrogen
envelope in solar units (bottom panel) as a function of WD age for the
0.33-\msun oxygen WD model. Solid lines correspond to the case
when element diffusion is considered and dashed lines when diffusion
is neglected. At each flash episode, the surface abundance of hydrogen
is abruptly reduced as a result of convective mixing. Afterwards,
diffusion causes the bulk of hydrogen to float on the surface again.
Note also that during each flash episode, the mass of the hydrogen
envelope is considerably reduced.  Models with diffusion enter the
final cooling branch (ages larger than $10^7$ yr) with a markedly
smaller hydrogen mass. By contrast models without diffusion retain a
larger hydrogen envelope which is largely burnt during the ensueing
evolutionary stages.}
\label{figxmh}
\end{figure}

The time dependence of the surface hydrogen abundance and the mass of
the hydrogen envelope for the evolutionary stages following the end of
the pre-WD evolution is illustrated in Fig. \ref{figxmh}. The role of
element diffusion is clearly emphasized by this figure.  During each
short-lived flash episode, convection reaches deep enough into the
star to dredge helium up, thus leading to envelopes made up by
hydrogen and helium. This is responsible for the steep declines of the
surface hydrogen abundance.  For models with diffusion, after each
flash, the purity of the outer layers is rapidly reestablished as soon
as the star returns onto the cooling branch. Note that the time
intervals during which the envelope remains helium-enriched are indeed
extremely short.  This is in contrast to the predictions of models
without diffusion. For these diffusion-free models, there is no
physical process other than convection that can transport hydrogen to
the surface.  Thus, the final hydrogen abundance ($X_H \approx 0.35$)
is established by the mixing episode that occurred during the last
flash.  The mass of the hydrogen envelope is shown in the bottom panel
of Fig.  \ref{figxmh}. An important feature revealed by this figure is
that the remaining amount of hydrogen, shortly after hydrogen flashes
have ceased, is markedly lower when diffusion is
considered. Specifically, the hydrogen mass at the beginning of the
final cooling track amounts to $M_H= 2.52 \times 10^{-4}$\msun, as
compared with the $M_H= 3.82 \times 10^{-4}$\msun that remains for the
case when diffusion is neglected. The smaller hydrogen masses left in
models with diffusion can be understood on the basis that, as the WD
evolves along the cooling branch, the tail of the hydrogen
distribution chemically diffuses inwards where the temperature is high
enough to burn it, a process which, as we mentioned, eventually
induces the occurrence of an additional thermonuclear flash. As can be
seen, during this last flash episode the total amount of hydrogen in
the star is noticeably reduced as a result of the nuclear burning. As
we shall see below, this fact produces a different cooling history
even at very late stages of evolution.

\begin{figure}
\includegraphics[width=8cm,height=8cm]{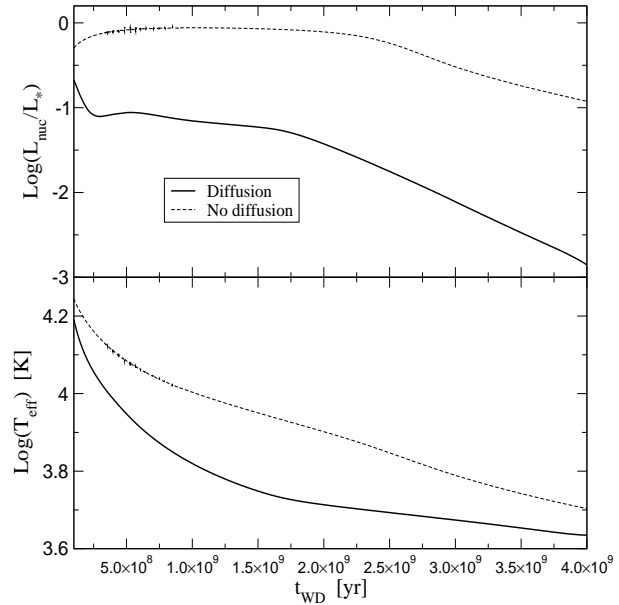}
\caption{Ratio of nuclear-to-surface luminosities (upper panel) and
effective temperature (bottom panel) as a function of WD age
for the 0.33-\msun oxygen WD model. Solid lines correspond to the case
when diffusion is considered and dashed lines when diffusion is 
neglected. Because the larger hydrogen envelope mass characterizing
models without diffusion (see figure \ref{figxmh}), hydrogen burning for
such models is the dominant source of energy even at advanced stages of
evolution, with the result that evolution is considerably slowed down.}
\label{figlteff}
\end{figure}

It is important to note that when diffusion is neglected, an
appreciable amount of hydrogen is processed {\it over the final
cooling phase}. Accordingly, we expect that the role played by stable
hydrogen shell burning during the final WD cooling phase is different
according to whether element diffusion is considered or not, an
expectation that is indeed borne out by the results shown in
Fig.~\ref{figlteff}. In the upper panel of this figure, we depict the 
ratio of nuclear-to-surface
luminosities as a function of WD age for the 0.33-\msun oxygen-core
WD. The solid line corresponds to the case when diffusion is included
and the dashed line to the situation without diffusion. Only the
evolutionary stages corresponding to the final cooling branch is
depiced in the figure.  Note that for these evolutionary stages of
star models with diffusion, nuclear energy release never constitutes
the main contribution to the radiated luminosity.  In particular, as
the WD evolves across the domain of the ZZ Ceti stars, nuclear burning
contributes at most 10 per cent to the surface luminosity.  It is
worth mentioning that the nuclear energy production for effective
temperatures lower than $\approx$ 11\,000K is almost entirely from the
proton-proton chain. In fact, hydrogen burning via CNO-cycle reactions
abruptly ceases by the time the WD has approached the hot edge of the
ZZ Ceti instability strip.  Eventually, hydrogen burning becomes
virtually extinct at the lowest luminosities that we computed. These
results are in contrast to the situation in which element diffusion
is neglected. Here, stable hydrogen burning via proton-proton chain
becomes the dominant energy source even at advanced stages of
evolution. Note that for this situation, hydrogen burning contributes
almost 80-90 per cent to surface luminosity even at $T_{\rm eff}
\approx$ 8000K.  Thus, we conclude that {\it element diffusion
prevents hydrogen burning from being a main energy source for most of
the evolution of a low-mass, oxygen-core WD}. This conclusion agrees
with the predictions of Althaus et al. (2001ab) for the case of
low-mass, helium-core WDs.

Not surprisingly, the above-mentioned results have strong implications
regarding the evolutionary ages of the models.  This is documented in
the bottom panel of Fig.  \ref{figlteff}, which illustrates the
effective temperature versus the age of the WD. Again, the solid line
corresponds to the evolutionary sequence with element diffusion.  Note
that when diffusion processes are considered, cooling ages become
roughly a factor 2-3 smaller.  In fact, because nuclear reactions in
models with diffusion do not dominate the energetics once the star has
settled upon its final cooling track, to maintain its luminosity, the
WD must extract energy from its relic thermal content, thus causing a
much faster cooling. On the contrary, in the case of models without
diffusion, evolution is dictated by residual hydrogen burning, giving
rise to very long cooling ages.  For instance, to reach $T_{\rm eff}=$
10\,000~K, models with diffusion require 0.36~Gyr, whilst models without
diffusion need about 1~Gyr.

\begin{figure}
\includegraphics[width=8cm,height=8cm]{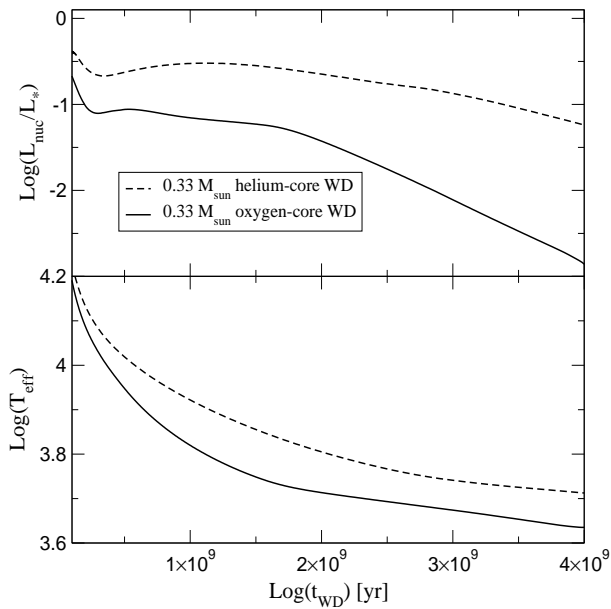}
\caption{Ratio of nuclear-to-surface luminosities (upper panel) and
effective temperature (bottom panel) as a function of WD age
for 0.33-\msun  WD sequences with diffusion. Solid (dashed) lines 
correspond to the case of oxygen-core (helium-core) models. For both cases,
hydrogen burning do not represent the main source
of energy. Note that oxygen-core WDs evolve much more rapidly than
their helium-core counterparts.}
\label{heox}
\end{figure}

We compare now the evolutionary properties of our 0.33-\msun
oxygen-core WD models with those of their helium-core
counterparts. Specifically, the comparison is made with a 0.327\msun-
helium WD sequence with diffusion as calculated by Althaus et
al. (2001a). We begin by examining the upper panel of Fig. \ref{heox}
which illustrates the role of hydrogen burning as energy source for
both sequences during the final cooling phase.  For the stages
depicted in the figure, nuclear burning is not the dominant source of
star's luminosity for both core compositions (see Althaus et al. 2001a
for the evolutionary properties of helium-core WDs when diffusion is 
considered); thus evolution is
basically governed by the decrease of the thermal heat content of the
star.  As is well known, for a given mass, an oxygen core has a
lower total capacity of storing heat than a helium core (we remind the
reader that the specific heat is proportional to the number of
particles; hence, per gram of material, the specific heat is inversely
proportional to the atomic weight of the constitutive elements).
Accordingly, we expect oxygen-core WDs to cool faster than helium-core
ones.  This expectation is confirmed by the bottom panel of
Fig.~\ref{heox} which shows the time taken by objects to cool down to
a given efective temperature.  We find that, at advanced stages, to
reach a given temperature, the oxygen-core WD has to evolve in about
half the time a helium-core WD need.

\begin{figure}
\includegraphics[width=8cm,height=8cm]{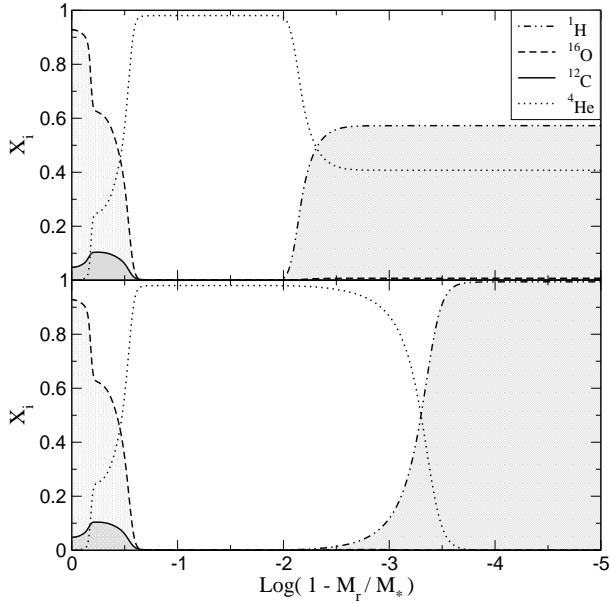}
\caption{Chemical abundance distribution in terms of the outer mass 
fraction for
the 0.33-\msun oxygen-core WD. Upper panel corresponds to an evolutionary
stage shortly after the formation of the oxygen core
and bottom panel to an evolutionary stage near the ZZ Ceti 
instability strip. Note that 
element diffusion markedly modifies the shape of the hydrogen and helium 
profiles.}
\label{inifin}
\end{figure}

The shape of the chemical composition profile is critical regarding
the pulsational properties of WDs.  In particular, it contributes to
the shape of the Ledoux term appearing in the Brunt-V\"ais\"al\"a
frequency (Brassard et al.  1991) and plays a critical role in the
phenomenon of mode trapping in WDs (see Brassard et al. 1992 and also
C\'orsico et al. 2001). In Fig.  \ref{inifin} we show the abundance by
mass of $^{1}$H, $^{4}$He, $^{12}$C and $^{16}$O as a function of the
outer mass fraction for the 0.33-\msun oxygen-core WD remnant. The
upper panel depicts the abundance distribution shortly after the formation
of the oxygen core and before the remnant experiences its first
thermonuclear flash.  The internal chemical profile emerging from the
pre-WD evolution is characterized by an almost pure oxygen core of
0.08\msun.  The oxygen core is surrounded by a shell rich in helium,
carbon, and oxygen with an overlying essentially pure helium buffer.
The abundance distribution in the innermost regions is clearly
different from that expected in the case of helium-core WDs, a feature
which is expected to bear its signature in the theoretical period
spectrum of these stars.  The outermost layers are composed by helium
and hydrogen.  The abundance of these elements is different from those
assumed for the interstellar medium because mass loss during the close
binary evolution exposed layers in which hydrogen burning has occurred
at earlier times. On the cooling track, the abundance distribution is
altered by gravitational settling and chemical diffusion\footnote{It
should be kept in mind that an appreciable fraction of the hydrogen
envelope is consumed by nuclear burning during the flash episodes.}.  
This is visualized
in the bottom panel of Fig.  \ref{inifin} that illustrates the
resulting chemical profile when the WD has reached the domain of the
ZZ Ceti instability strip on the final cooling branch. Note that
element diffusion appreciably modifies the abundance distribution,
particularly in the outer layers. Indeed, the
hydrogen and helium profiles are markedly smoothed out by the time the
ZZ Ceti domain is reached. At deeper layers, diffusion time-scale
becomes much longer than the evolutionary time-scale and the chemical
profile remains thus fixed during the entire WD evolution.

\begin{figure*}
\includegraphics[width=15cm,height=15cm]{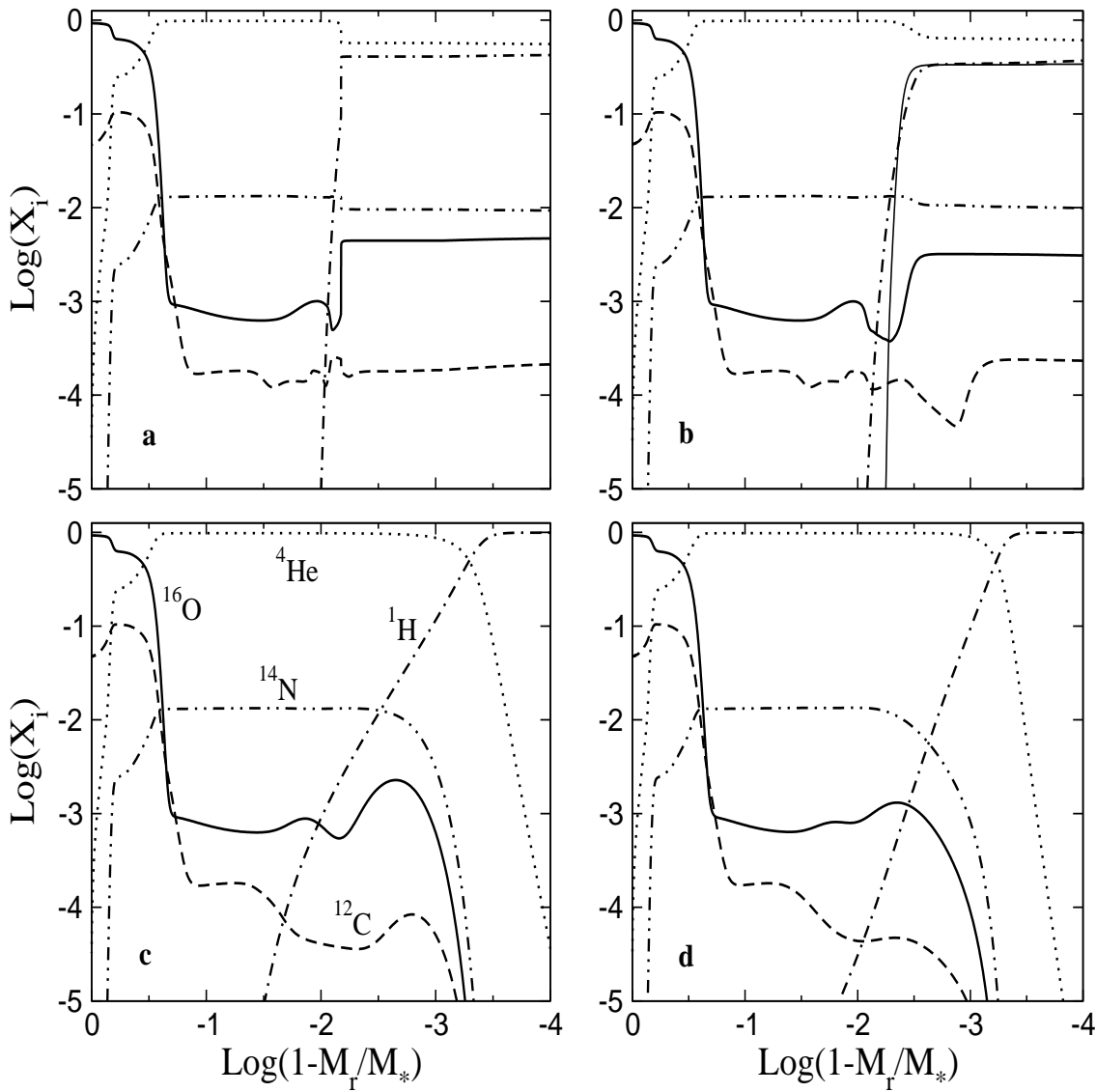}
\caption{Chemical abundance distribution for $^1$H, $^4$He, $^{12}$C,
$^{14}$N and $^{16}$O as a function of the outer mass fraction for the
0.33-\msun oxygen-core WD at four selected evolutionary stages.  Panel
{\bf a} shows the chemical stratification during the second flash
episode, shortly after the occurrence of the outer convective
mixing. Panel {\bf b} corresponds to the stage just before the
occurrence of the last flash. To emphasize the importance of chemical
diffusion, we also show with thin solid line the hydrogen profile
emerging from the previous flash. Panels {\bf c} and {\bf d}
correspond to models on the final cooling branch: the ZZ Ceti domain
and the last computed model, respectively. Note the role of both
gravitational settling and chemical diffusion. In particular, chemical
diffusion causes some hydrogen to reach deeper layers. At high $T_{\rm
eff}$ values (panel {\bf b}) this effect is responsible for the
occurrence of the last flash episode. Values of (log(${L/\rm
L_{\odot}})$ and log$T_{\rm eff}$) for panels {\bf a}, {\bf b}, {\bf
c} and {\bf d} are, respectively: (0.96; 4.34), (0.72; 4.64), (-2.53;
3.98) and (-4.03; 3.63). }
\label{quimlog}
\end{figure*}

The effect of diffusion on the element distribution within the star is
emphasized in Fig.~\ref{quimlog}, where the abundances of $^1$H,
$^4$He, $^{12}$C, $^{14}$N and $^{16}$O are shown as a function of the
outer mass fraction for 0.33-\msun oxygen-core WD models at four
selected epochs.  The effect of diffusion on the element distribution
is clearly noticeable. In particular, chemical diffusion moves a
hydrogen tail inwards into hotter layers, causing the hydrogen burning
rate to increase. At high $T_{\rm eff}$ values (panel {\bf b}) this
effect is responsible for the occurrence of the last flash episode. It
is interesting to note that by the time the ZZ Ceti instability strip
is reached (panel {\bf c}), there exists an appreciable tail of hydrogen 
in the helium buffer.  In fact, by this time the hydrogen distribution 
reaches its maximum depth.  At the same time, there are no metals in the 
outermost $2 \times 10^{-4}$ \msun of the star due to the occurrence of
gravitational settling.

\section{Pulsational analysis} \label{sec:results}

We have seen that the evolutionary properties of the oxygen-core WDs
differ appreciably from those of helium-core WDs, a fact which could
eventually be used to shed light on the internal composition of
observed low-mass WDs. However, a more promising and potential way of
distinguishing both types of stars would be by means of the study of
their pulsational patterns if low-mass WDs were in fact found to
pulsate as ZZ-Ceti type variables.  Although signs of variability
in low-mass WDs have not been detected so far, we judge that an
exploration of their oscillatory properties is nevertheless
worthwhile.  Indeed, in the context of more massive WDs, it is well
known that asteroseismology is a very powerful tool for probing into
the internal composition of degenerate stars (see, e.g., Bradley 1998 
and Metcalfe 2003). In particular, we are interested here in studying 
both the adiabatic and non-adiabatic pulsational properties.

\subsection{Adiabatic pulsations}

In order to explore the adiabatic pulsational properties of the WD models 
presented in this work, we have employed a general Newton-Raphson 
pulsational code coupled to the LPCODE evolutionary code as detailed in 
C\'orsico et al. (2001, 2002; see also C\'orsico 2003). In particular, 
the computation of 
the Brunt-V\"ais\"al\"a frequency ($N$), an issue of fundamental relevance 
in the context of pulsating WD, is performed by adopting the prescription 
suggested by Brassard et al. (1991). Specifically, our code computes the 
so called ``Ledoux term'' ($B$), given by

\begin{equation}
B=-\frac{1}{\chi_{_{\rm  T}}}  \sum^{n-1}_{{\rm  i}=1}  \chi_{_{X_{\rm
i}}}
\frac{d\ln {X}_{\rm i}}{d\ln P},
\end{equation}

\noindent with  

\begin{equation} 
\chi_{_{X_{\rm i}}}= \left( \frac{\partial \ln{P}}
{\partial \ln{X_{\rm i}}} \right)_{\rho,T,\{X_{\rm j \neq i}\} },
\end{equation}

\noindent and 

\begin{equation}
\chi_{_{\rm  T}}= \left( \frac{\partial \ln P}{\partial \ln T}
\right)_{\rho, \{X_i\}}.
\end{equation}

Any local feature in the profiles of the internal chemical abundances 
is reflected by the Brunt-V\"ais\"al\"a 
frequency through the Ledoux term, as it can be clearly noted from the 
following expression for $N$ (Brassard et al. 1991):

\begin{equation} 
N^2   =   \frac{g^2\   \rho}{P}\   \frac{\chi_{_{\rm  T}}}{\chi_{\rho}}\
\left(\nabla_{\rm ad} - \nabla + B \right).
\end{equation}

\noindent Here, $X_i$ is the mass fraction of specie $i$, $n$ the number of 
ionic species present in the plasma, $g$ the local gravity, 
$\nabla_{\rm ad}$ and $\nabla$ the adiabatic and actual 
temperature gradients, respectively, and 

\begin{equation}
\chi_{_{\rm  \rho}}= \left( \frac{\partial \ln P}{\partial \ln \rho}
\right)_{T, \{X_i\}}.
\end{equation}
  
\begin{figure}
\includegraphics[width=85mm,height=85mm]{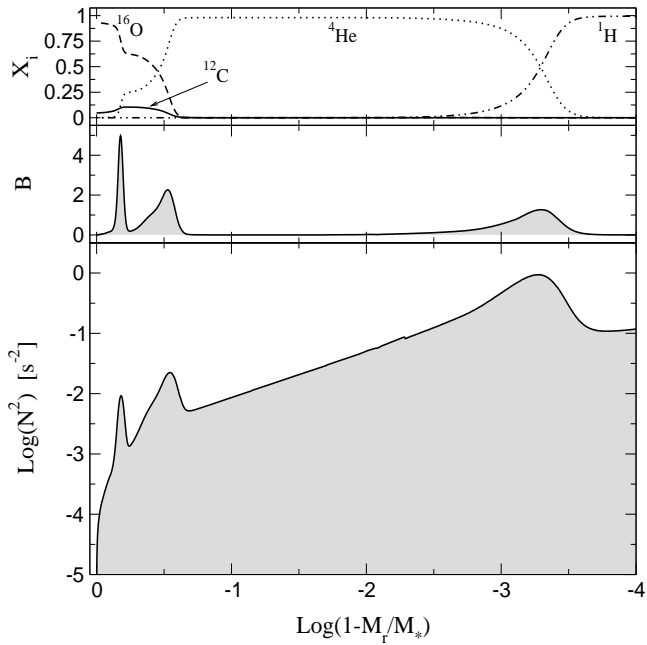}
\caption{The internal chemical profile, the Ledoux term and the 
Brunt-V\"ais\"al\"a frequency (upper, middle and bottom panel, 
respectively) in terms of the outer mass fraction, corresponding to 
a selected 0.33-\msun oxygen-core WD model at $T_{\rm eff} \approx$ 10000 K.
Note that $N$ exhibits a peaked feature  at the innermost region of 
the model. It is a direct consequence of  the steep change  of the $^{16}$O 
abundance at $\log(1-M_r/M_*) \approx -0.2$. The remainder, less narrow 
features, are associated with the oxygen-helium and the helium-hydrogen 
interfaces.}
\label{xbn2co}
\end{figure}

\begin{figure}
\includegraphics[width=85mm,height=85mm]{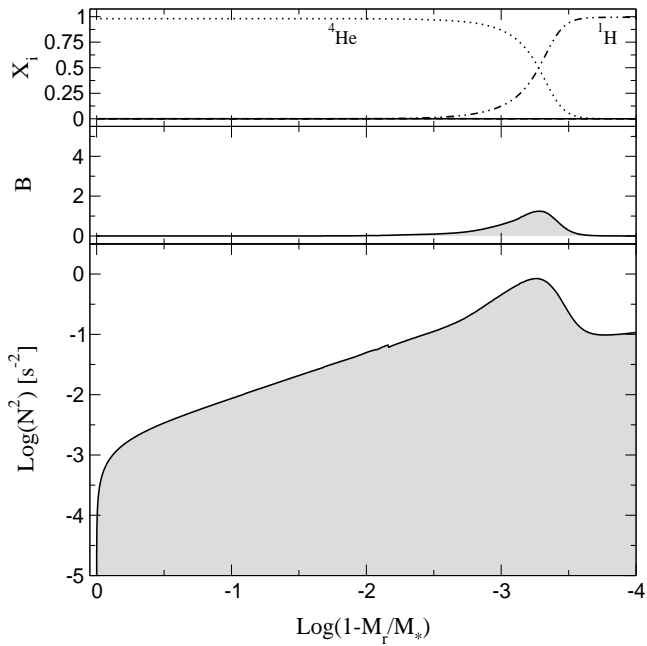}
\caption{Same as Fig. \ref{xbn2co}, but for a helium-core WD model of 
0.33-\msun at the same effective temperature. In this case, the only 
feature characterizing the shape of the Brunt-V\"ais\"al\"a frequency 
is that associated with the helium-hydrogen interface (the only chemical 
transition region in this model).}
\label{xbn2he}
\end{figure}

In this section we explore the adiabatic pulsational
properties of low-mass, oxygen-core WD models and we compare them with
those corresponding to the helium-core counterparts.  Specifically, we
have computed the adabatic pulsational pattern of gravity modes
corresponding to an evolutionary sequence of 0.33-\msun oxygen-core WD
models for an effective temperature range of 15\,000-8000~K.
Additionally, we have carried out pulsational calculations
corresponding to an evolutionary sequence of 0.33-\msun helium-core WD
models for the same $T_{\rm eff}$ interval.  To assess the
mode-trapping/confinement features associated with both kind of
objects, we restricted ourselves to analyzing two selected models
picked out from both sequences at $T_{\rm eff} \approx 10\,000$
K. Hereafter, we shall denote these template models as OCWD
(oxygen-core WD) and HECWD (helium-core WD) models.  We
begin by examining the shape of the Ledoux term and the
Brunt-V\"ais\"al\"a frequency in Figs. \ref{xbn2co} and \ref{xbn2he}
(middle and bottom panels, respectively). To make easier the
association amongst the various chemical transitions and the
corresponding features in $N$, we have also included in these figures
the internal chemical profiles (upper panels).

For the case of the OCWD model, the most outstanding characteristic in the
shape of $N$ is the peaked feature at $\log(1-M_r/M_*) \approx -0.2$
(Fig. \ref{xbn2co}). This is directly associated with the abrupt
variation in the oxygen abundance at such very deep layers. On the
other hand, the oxygen-helium interface is responsible for the less
narrow neighbouring peak in $N$, and finally, at $\log(1-M_r/M_*)
\approx -3.25$ there is a broad, smooth bump which is caused by the
helium-hydrogen transition region. The shape of $N$ corresponding to
the HECWD model is documented in the bottom panel of
Fig. \ref{xbn2he}. As can be appreciated from the figure, in this case
the only local feature present in $N$ is that resulting from the
helium-hydrogen chemical transition zone.  The comparison of the
Brunt-V\"ais\"al\"a frequency for both models deserves some comments.
As is well known, any local feature in the Brunt-V\"ais\"al\"a
frequency constitutes a source for trapping or confinement of modes
(see Brassard et al. 1992; C\'orsico et al 2002). As a result, we
expect the mode-trapping/confinement properties of OCWD model to be
considerably more featured than those of the HECWD model. The other
point worthy of comment is the fact that $N$ in the HECWD model shows
slightly greater values at the innermost core region than for the OCWD
case. Although this seems to be a subtle difference, it causes the
periods to be shorter and the asymptotic period spacing 
to be somewhat smaller in the case of the OCWD model.

In what follows we comment on the results we obtained for the
pulsational properties of our models.  Specifically we have computed
the adiabatic, non-radial $g$-modes of low degree ($\ell= 1, 2, 3$)
with overtones covering a wide period window of $50-2000$ sec. In
addition, related quantities such as oscillation kinetic energy
($E_{\rm{kin}}$) and first-order rotational splitting coefficient
($C_{\ell,k}$) have been obtained for each computed mode. 
We mention that the radial eigenfunction ($\delta r/r$) is normalized
to 1 at the surface. The
main observable quantities in pulsating WD stars are the periods ($P_k$) 
and the forward period spacing ($\Delta P_k= P_{k+1}-P_k$).  Thus, we 
shall employ these quantities as the main discriminants to demonstrate 
the different pulsational behaviours associated with OCWD and HECWD 
model stars.

\begin{figure}
\includegraphics[width=85mm,height=85mm]{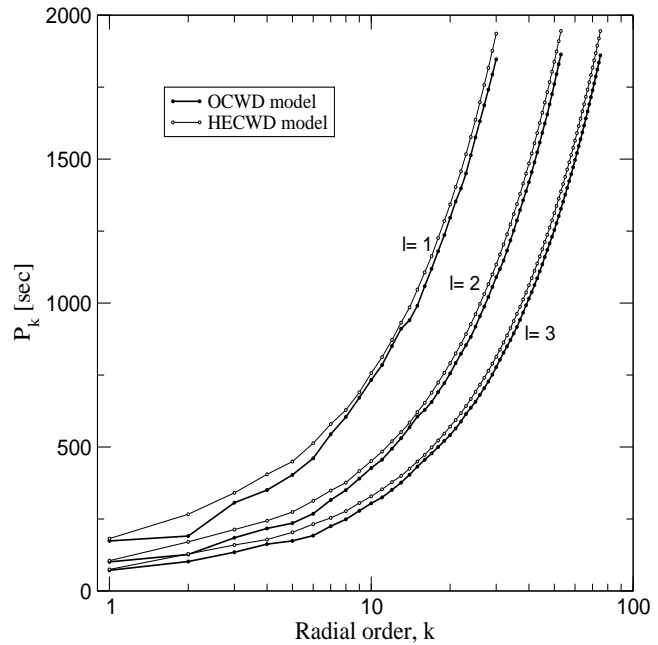}
\caption{Periods as a function of the overtone number for 
$\ell$ =1, 2 and 3 corresponding to OCWD (thick lines, fill dots) and 
HECWD (thin lines, hollow dots) models.}
\label{periods}
\end{figure}

We begin by examining Fig.  \ref{periods} which shows the periods as a
function of the  overtone number for $\ell$ =1,  2 and 3 corresponding
to OCWD  (thick lines) and HECWD  (thin lines) models.  Note that {\it
all} the periods corresponding to  HECWD model are longer than for the
case of the OCWD one. As mentioned, this is a consequence of $N$ being
slightly larger at the innermost core  region in the case of the HECWD
model.  Note also for  this model  the near  uniformity in  the period
distribution as compared with the OCWD case (see below).

\begin{figure*}
\includegraphics[width=15cm,height=15cm]{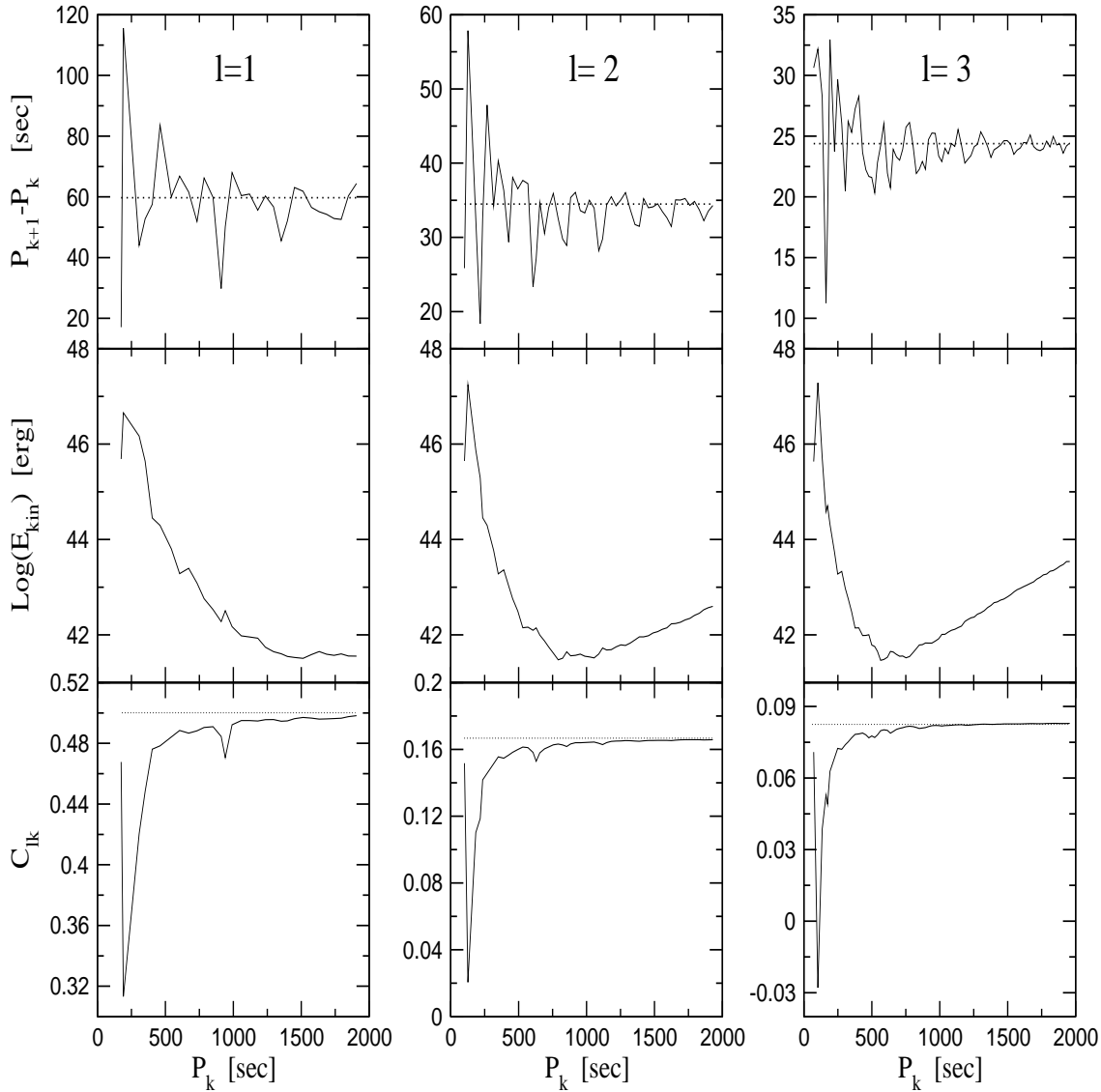}
\caption{The forward period spacing (upper panels), the logarithm 
of the kinetic energy (middle panels) and the first order rotational 
splitting coeficients (lower panels), in terms of the periods for $g$-modes 
with $\ell=1$ (left), $\ell=2$ (centre) and $\ell=3$ (right), for the same 
WD model analyzed in Fig. \ref{xbn2co} (OCWD models). 
In the interest of clarity we have 
ommited the symbols representing pulsation modes. Dotted lines represent 
the asymptotic period spacing (upper panels) and the asymptotic value of 
$C_{\ell,k}$ (lower panels).}
\label{pulsa-co}
\end{figure*}

\begin{figure*}
\includegraphics[width=15cm,height=15cm]{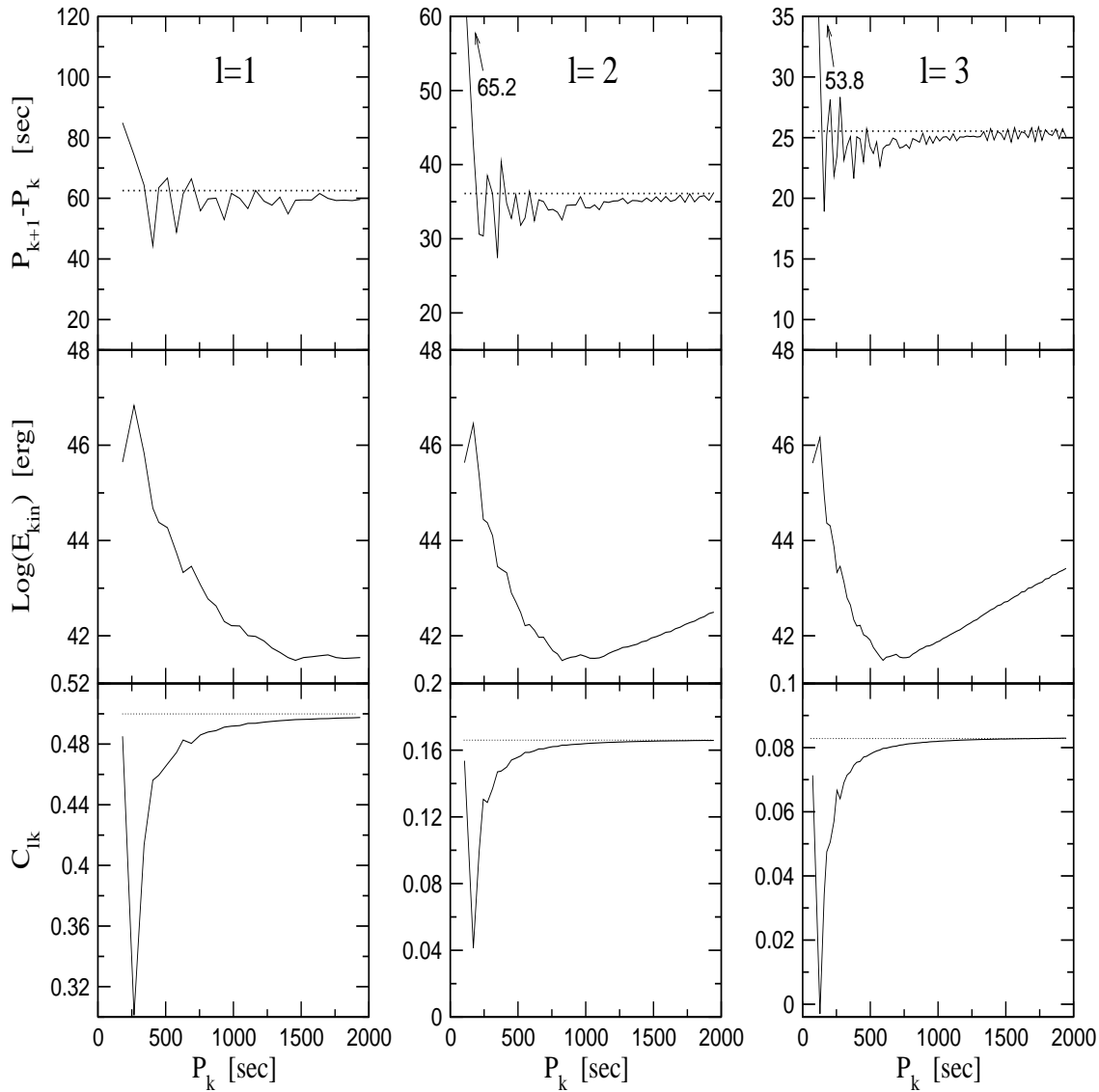}
\caption{Same as Fig. \ref{pulsa-co} but for the model analyzed in 
Fig.\ref{xbn2he} (HECWD model).}
\label{pulsa-he}
\end{figure*}

In Fig.~\ref{pulsa-co} we show $\Delta P_k$, $\log(E_{\rm{kin}})$ and
$C_{\ell,k}$ (upper, middle and bottom panels, respectively) in terms
of periods, for $\ell= 1, 2, 3$ (left, centre and right)
corresponding to the OCWD model. Fig.~\ref{pulsa-he} depicts the same
quantities but for the case of the HECWD model. We begin by examining
the $\Delta P_k$ distribution in Fig.~\ref{pulsa-co}. As expected, the
spacing of consecutive periods is markedly featured, reflecting the
three sources of trapping/confinement of modes corresponding to each
of the three chemical interface.  An inspection of the figures reveals
that the amplitude of $\Delta P_k$ corresponding to the OCWD model is
typically larger as compared with the HECWD model. Note also the
strong minima in $\Delta P_k$ in the OCWD model. These minima
correspond to modes {\it partially confined} to the deepest regions of
the star. Indeed, at the high-density zone bounded by the stellar
centre and $\log(1-M_r/M_*) \approx -0.2$, such modes have larger
amplitudes of their eigenfunctions as compared with those of the
neighboring overtones. Consequently, such modes must exhibit local maxima 
in the $E_{\rm kin}$ distribution (see middle panels of
Fig.~\ref{pulsa-co}).  In contrast, the much less pronounced minima
exhibited by $\Delta P_k$ in the HECWD model are related to trapped
modes in the outer layers.  Thus, these modes are characterized by low
kinetic energy values (Fig.~\ref{pulsa-he}). Finally, we note that the 
distribution of the $C_{\ell,k}$ values corresponding to the HECWD model
exhibits less structure as compared with the case of the  OCWD one, 
particularly regarding to the short- and intermediate-period domains.    

Clearly, from the above considerations we conclude that there exist marked 
differences beetwen the adiabatic pulsational properties of helium- and 
oxygen-core low-mass WDs. This in principle turns the 
asteroseismological techniques usually employed in variable WD studies  
into an important tool for distinguishing both types of stars. 

\subsection{Non-adiabatic pulsations}

\begin{figure}
\includegraphics[width=8cm,height=8cm]{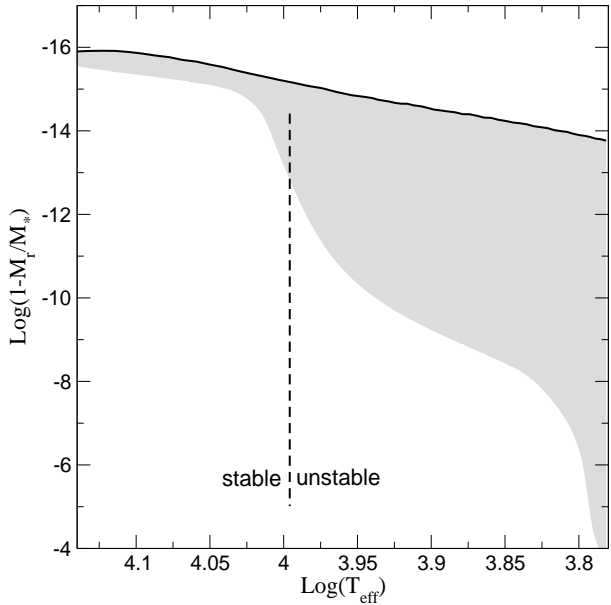}
\caption{Location of the convection zone as given by the outer mass 
fraction as a function of effective temperature (grey domain). The vertical,
dashed line indicates the value of effective temperature below which 
$g$-modes become unstable. The thick line marks the outer edge of the 
convection zone which is near the location of the photosphere.}
\label{conve}
\end{figure}

Finally, we comment on the main results from non-adiabatic pulsation
calculations of the oxygen-core WD models. We emphasize that for the
WD regime, we modelled convection according to the mixing-length theory
setting the mixing length to $1.5$ pressure scale-heights. The grey
domain in Fig.~\ref{conve} shows the extension of the superficial
convection zone as a function of effective temperature.  Note that at
$T_{\rm eff} \approx 10\,000$ K, below which modes are found unstable
(see below), the base of the convection zone sinks appreciably into
the star. The linear non-adiabatic oscillation spectra were computed
for dipolar ($\ell=1 $) $g$-modes in the period range between 100 and
about 1200~sec using the same Riccati shooting-method as referred to
in Gautschy et~al. (1996) (see also Gautschy \& Althaus 2002).  The
non-adiabatic computations were done assuming a frozen-in convective
flux.

\begin{figure}
\hskip -5mm
\includegraphics[width=9cm]{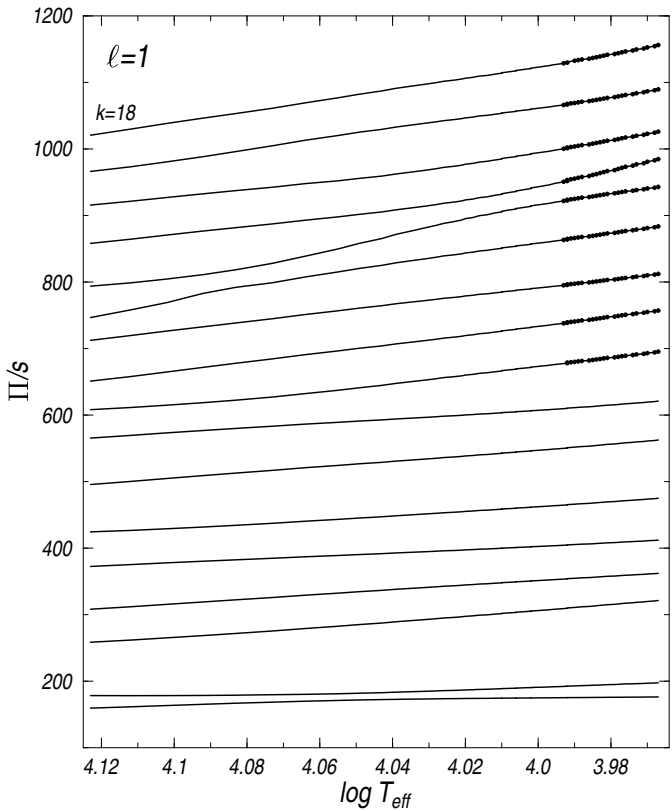}
\caption{Modal diagram for $\ell= 1$ of the 0.33-\msun oxygen-core white 
dwarf sequence. The dots on some intermediate-order modes indicate 
their instability.}
\label{modal}
\end{figure}

Figure~\ref{modal} displays the modal diagram for the 0.33-\msun
oxygen-core WD sequence. Specifically, the figure depicts the periods
of the lowest order $\ell = 1$ modes as a function of effective
temperature. Overstable modes are indicated by black dots. Like radial
orders are connected by a continuous line. The highest order shown is
$k = 18$.  An instability domain is found that starts at $\log T_{\rm
eff}= 3.993$ and continues below the cool end of the model series
analyzed. Only modes with periods exceeding about 680~s become
unstable. A striking feature concerns the position of the blue edge:
It does not depend on the radial order, i.e. on the length of the
periods. A survey computation showed furthermore that the saturation
of the excitation must be beyond $k=30$.  In other words, we did not
find the long-period edge of the instability domain.
    
\begin{figure}
\hskip -5mm
\includegraphics[width=9cm]{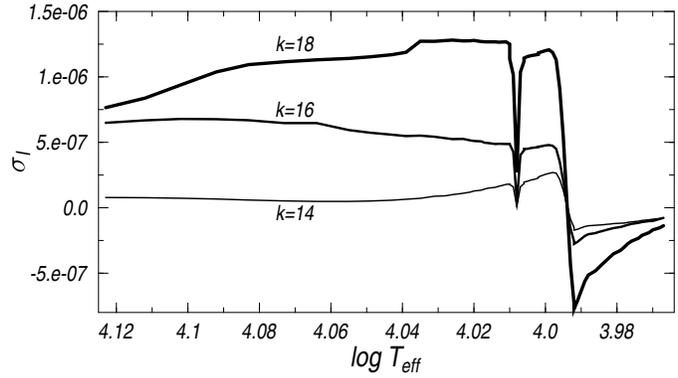}
\caption{Same of Figure \ref{modal}, but for the imaginary part of
selected eigenfrequencies. Negative numbers mean oscillatory instability. 
The radial order of the eigenmodes is labelled on the graph.}
\label{imag}
\end{figure}

Figure \ref{imag} shows the imaginary parts of the eigenvalues as a
function of $T_{\rm eff}$. Note that the maximum of instability,
i.e. the minimum of $\sigma_{\rm{I}}$, is encountered very close to
the blue edge. The excitation rates decline quickly toward lower
temperatures. Notice, as referred to in the last paragraph, that the
instability (i.e., the magnitude of $\sigma_{\rm I}$) continues to get
stronger to higher overtone numbers.

We mention that the  non-adiabatic computations included the effect of
nuclear burning, i.e. the $\epsilon$-mechanism. However, we considered
only  the equilibrium  situation, neglecting  the perturbation  of the
abundance of  the reactants. We find  that the work  integrals have no
appreciable  contribution   from  $\epsilon$-mechanism.  A   point  of
interest  is the  question if  pulsations  might trigger  a very  late
hydrogen shell flash.  Based on the fact that  the eigenfunctions have
very low  amplitudes in hydrogen-burning  regions, we do not  expect a
pulsation-induced thermonuclear flash to be feasible\footnote{The $g$-mode 
pulsations' preponderant horizontal displacements are unlikely
to disturb the geometry of the nondegenerate H-burning shell such as to
trigger a thermal runaway.}.

Finally, since no convection~--~pulsation interaction was considered, the
non-adiabatic results indicate at the very best that the $\kappa$
mechanism destabilizes low-degree modes of intermediate order. 
The action of convection can still modify the picture considerably,
even on a qualitative level.

\section{Conclusions} \label{sec:conclusion}

Motivated by theoretical evidence suggesting that some of the
presumed low-mass, helium-core WDs could actually be WDs with oxygen
cores (Iben \& Tutukov 1985 and Han, Tout \& Eggleton 2000), we explored 
in this paper the
evolution of low-mass WDs with oxygen cores in a
self-consistent way with the predictions of nuclear burning, the
chemical evolution caused by time-dependent element diffusion, and the
history of the white dwarf progenitor. We concentrated on an
oxygen-core WD remnant of mass 0.33 \msun.  The evolutionary
stages prior to the white dwarf formation were also fully accounted
for by computing the conservative binary evolution of an initially
2.5-\msun Pop.~I star with a  1.25 \msun companion, and an
orbital period of $P_i$= 3 days.  In addition, we explored the
adiabatic and non-adiabatic pulsational properties of the resulting WD
models.

As for the main evolutionary results we mention:

- The star's evolution is characterized by the occurrence of several
episodes of thermonuclear flashes due to unstable hydrogen burning at
the base of the hydrogen-rich envelope.

- Diffusion processes play an important role in the evolution of
low-mass oxygen-core WDs. In particular, the remaining amount of
hydrogen, shortly after hydrogen flashes have ceased, is markedly
lower when diffusion is considered. This fact produces a different
cooling history even at very late stages of the evolution. In
particular, element diffusion prevents hydrogen burning from being the
main energy source for most of the evolution of a low-mass,
oxygen-core WD. As a result, cooling ages are roughly a factor
$2~-~3$ smaller than ages derived from models neglecting diffusion.

- As compared with the helium-core counterparts, low-mass, oxygen-core 
WDs cool much faster once they have settled on their final cooling track.  
In particular, at advanced stages, to reach a given temperature, the 
oxygen-core WD evolves about twice as fast as a helium-core WD.

The main  pulsational properties of our models are:

- The structure of the pulsational spectrum of the oxygen-core WD
models is notably richer than that of the helium-core ones. This fact
is apparent from the period spacing diagrams, which exhibit large
amplitudes of trapping/confinement imprints.
 
- The presence of an oxygen core is responsible for some {\it
partially confined} modes that are characterized by comparatively 
large values of kinetic energy and strong minima in a period-spacing 
diagram.

- Modes with periods longer than about 680 sec become unstable at 
effective temperatures below about $10000$~K. 

We conclude that both the pulsational and evolutionary properties of
low-mass, carbon-oxygen WDs turn out to be sustantially different from
those of the helium-core WDs. In particular, our results suggest that
a discrimination between helium-core and carbon/oxygen-core WDs by
means of asteroseismology is possible if low-mass WDs were in fact
found to pulsate as ZZ Ceti~--~type variables.  Despite the lack of
signs of variability in low-mass WDs so far, we consider the
monitoring of low-mass WDs to find variable candidates to be
worthwhile.

Complete tabulations of the results of the present paper are available
at http://www.fcaglp.unlp.edu.ar/evolgroup/

We warmly acknowledge to our referee Michael Montgomery for his
suggestions and comments that improved the original version of this
work. We also thank the financial support by the Instituto de 
Astrof\'{\i}sica de La Plata. This research has made use of NASA's 
Astrophysical Data System Abstract Service.

\bsp

\label{lastpage}

\end{document}